\documentclass[
amsmath, %
floatfix, %
twocolumn, %
reprint, %
prx, %
aps, %
citeaut itoscript, %
longbibliography, %
final, %
]{revtex4-2}
\renewcommand{\thispagestyle}[1]{}
\usepackage[]{newtxtext} 
\usepackage[subscriptcorrection,nosymbolsc,smallerops,bigdelims]{newtxmath}
\DeclareMathAlphabet{\mathcal}{OMS}{cmsy}{m}{n}
\DeclareMathAlphabet{\mathbcal}{OMS}{cmsy}{b}{n}
\usepackage{bm}

\usepackage[low-sup]{subdepth}

\usepackage{comment}
\usepackage[utf8]{inputenc}
\usepackage[T1]{fontenc}

\usepackage[]{graphicx}
\usepackage{latexsym}
\usepackage{color}
\usepackage{mathtools}
\usepackage[section]{placeins}
\usepackage[]{xcolor}
\usepackage{bm}
\usepackage{bbold}
\usepackage{multirow}
\usepackage{siunitx}
\usepackage{hyperref}
\hypersetup{
	colorlinks,
	linkcolor={blue!90!black},
	citecolor={blue!90!black},
	urlcolor=	{blue!90!black}
}
\let\oldeqref\eqref
\renewcommand*{\eqref}[1]{%
	\hyperref[#1]{\oldeqref{#1}}%
}

\usepackage{floatrow}
\usepackage{placeins}

\usepackage{tikz}    

\usetikzlibrary{arrows.meta, decorations.pathmorphing, decorations.markings, calc}

\newcommand{\eps}{\hat{\varepsilon}}
\newcommand{\lck}{\lambda, k}
\newcommand{\lk}{\lambda k}
\newcommand{\lnk}{\lambda -k}
\newcommand{\vlck}{\lambda, \vk}
\newcommand{\lcnk}{\lambda, -k}
\newcommand{\vlk}{\lambda \vk}
\newcommand{\vk}{\bm{k}}
\newcommand{\vB}{\bm{B}}

\newcommand{\omz}{\omega_{\mr{Z}}}

\newcommand{\mr}[1]{\mathrm{#1}}

\newcommand{\rr}{\bm{r}}

\newcommand{\kp}{{\bm{k} {\cdot} \bm{p}}}

\mathchardef\mhyphen="2D
\newcommand{\upa}{{\uparrow}}
\newcommand{\doa}{{\downarrow}}

\DeclarePairedDelimiterX{\comm}[2]{\lbrack}{\rbrack}{#1, #2}

\DeclarePairedDelimiterX{\braket}[2]{\langle}{\rangle}{#1\delimsize\vert #2}
\DeclarePairedDelimiterX{\ketbra}[2]{\rvert}{\lvert}{#1 \delimsize\rangle\!\delimsize\langle #2}
\DeclarePairedDelimiterX{\matrixel}[3]{\langle}{\rangle}{#1 \delimsize\vert #2 \delimsize\vert #3}

\newcommand{\hc}{\mr{H.c.}}

\newcommand{\eqnref}[1]{Eq.~\eqref{#1}}
\newcommand{\tabref}[1]{Tab.~\ref{#1}}
\newcommand{\figref}[1]{Fig.~\ref{#1}}
\newcommand{\subfigref}[2]{Fig.~\hyperref[#1]{\ref*{#1}(#2)}}
\newcommand{\subfigsref}[3]{Figs.~\hyperref[#1]{\ref*{#1}(#2)}-\hyperref[#1]{\ref*{#1}(#3)}}
\newcommand{\aeqnref}[1]{Equation~\eqref{#1}}

\newcommand{\afigref}[1]{Figure~\ref{#1}}
\newcommand{\asubfigref}[2]{Figure~\hyperref[#1]{\ref*{#1}(#2)}}
\newcommand{\asubfigsref}[3]{Figures~\hyperref[#1]{\ref*{#1}(#2)}-\hyperref[#1]{\ref*{#1}(#3)}}

\usepackage[low-sup]{subdepth}
\usepackage{ragged2e}

\definecolor{cbred}{HTML}{e31a1c}
\definecolor{cbgreen}{HTML}{33a02c}
\definecolor{cbblue}{HTML}{176aa7}
\definecolor{cborange}{HTML}{ff7f00}
\definecolor{cbviolet}{HTML}{6a3d9a}
\definecolor{cbbrown}{HTML}{b15928}

\definecolor{cblred}{HTML}{fb9a99}
\definecolor{cblgreen}{HTML}{b2df8a}
\definecolor{cblblue}{HTML}{a6cee3}
\definecolor{cblorange}{HTML}{fdbf6f}
\definecolor{cblviolet}{HTML}{cab2d6}
\definecolor{cblbrown}{HTML}{ffff99}

\makeatletter
\newcommand{\doublewidetilde}[1]{%
  \mathpalette\doublewidetilde@{#1}%
}
\newcommand{\doublewidetilde@}[2]{%
  \ooalign{%
    $\m@th#1\widetilde{#2}$\cr
    \raise.35ex\hbox{$\m@th#1\kern1pt\widetilde{\phantom{#2}}$}\cr
  }%
}
\makeatother

\usepackage[activate={true,nocompatibility},final,tracking=alltext,kerning=true,spacing=true,protrusion=true,factor=1080,stretch=5,shrink=9,selected=true ,letterspace=-0]{microtype}
\usepackage{orcidlink}

\begin{document}

\title{Theory of phonon-induced spin relaxation in a structured phononic reservoir}

\author{Raseeb F. Haroon\,\orcidlink{0009-0000-9917-5182}}
\affiliation{Institute of Theoretical Physics, Wroc\l{}aw University of Science and Technology, 50-370 Wroc\l{}aw, Poland}

\author{Pawe{\l} Machnikowski\,\orcidlink{0000-0003-0349-1725}}
\affiliation{Institute of Theoretical Physics, Wroc\l{}aw University of Science and Technology, 50-370 Wroc\l{}aw, Poland}

\begin{abstract}
By combining Markovian and non-Markovian open quantum system theory with finite-element simulations, we develop a theory of electron spin relaxation in a structured phononic reservoir. This problem is crucial for understanding spin dynamics in hybrid systems involving mechanical modes, as well as for the design of devices combining spin degrees of freedom with photonic and phononic architectures, where the phonon density of states is modulated in the relevant spectral range corresponding to moderate magnetic fields. Taking a QD in a phononic waveguide as a representative and technologically relevant example, we show that spin relaxation in such environments is much more complex than in bulk. While the relaxation rates are several times to an order of magnitude higher than in bulk, there are parameter windows where the relaxation is suppressed by many orders of magnitude due to gaps in mode dispersion and selection rules imposed by mode symmetry. At the border between these two sectors, van Hove singularities in phonon dispersion lead to singularities in relaxation rates, for which we develop power-law scaling and develop a non-Markovian description of the dynamics, revealing polaronic dressing of the spin into slow acoustic modes. 
\end{abstract}

\maketitle

\section{Introduction}

Electron spins confined in self-assembled semiconductor quantum dots (QDs) are a leading solid-state platform for spin qubits in quantum information processing \cite{awschalom2013}. Self-assembled QDs combine optical preparation and readout, electrical control, and millisecond-scale spin lifetimes at low temperature \cite{kroutvar2004}. Spins confined in QDs can be interfaced \cite{Yilmaz2010} or entangled \cite{Delley2017} with photons, leading to efficient generation of indistinguishable photons \cite{Scholl2019}, which paves the way for solid-state-based quantum technologies, including cluster-state generation \cite{Lindner2009, Cogan2023a, Coste2023, Huet2025,Meng2024,schwartz2016}. 

The functionality of solid-state quantum systems can be greatly enhanced by coupling them to mechanical waves in the solid. This has allowed one to control the light-matter interaction of a self-assembled QD in the spectral \cite{metcalfe2010,weiss2021_optica} and temporal \cite{Wigger_2021_Spectral} domains. Acoustic couplings also provide a control mechanism for spin states of defect centers \cite{golter2016prx,whiteley2019,maity2020} and can mediate interaction with superconducting qubits \cite{Gustafsson_2014_Phonons,satzinger2018,Bienfait_2019_Entanglement}. Further enhancement of the efficiency and degree of control of the acoustic couplings is achieved by exploiting guided waves \cite{krenner2026_roadmap}. For instance, one can use a phononic crystal geometry, in which periodic patterning of a slab opens elastic band gaps, while a linear ``defect'' structure gives rise to one-dimensionally confined (guided) modes with frequency within this gap \cite{eichenfield2009,Kurosu2018,Hatanaka2020,rosinski_2026,Hatanaka2014}. Structures of this kind have been used to couple coherent acoustic fields to optically active quantum systems \cite{weiss2018,schuetz2015,Spinnler2024}, to drive NV centers in diamond and divacancy spins in silicon carbide \cite{golter2016,golter2016prx,whiteley2019}, and to bring individual mechanical modes close to their quantum ground state \cite{chan2011,Diamandi2025}.

While long spin lifetimes make QDs attractive for spin-based hybrid acousto-optical applications, they remain finite, setting a limit on spin storage, transduction, and coherent operations. At low temperature and in a moderate or strong magnetic field, the dominant relaxation channel for an electron spin in a III--V QD is spin--orbit- and phonon-mediated \cite{elzerman2004,amasha2008,meunier2007,Camenzind2018}. Here, the spin--orbit coupling admixes the two Zeeman sublevels, and the strain field of acoustic phonons drives transitions between the resulting hybrid states through the deformation-potential and piezoelectric interactions \cite{khaetskii2001,woods2002,golovach2004,bahder1990,mielnik-pyszczorski_2018}. 
The foundational treatments \cite{khaetskii2001,golovach2004,stavrou2017,woods2002} take the phonon environment to be a bulk continuum. In that case, the resulting spin-flip rate scales as a smooth power-law function of the magnetic field \cite{khaetskii2001,woods2002,golovach2004,mielnik-pyszczorski_2018}.  The value of this power-law exponent is set by the long-wavelength form of the spin-phonon coupling, which is constrained by the symmetry of the coupling, by the global charge conservation, and by the momentum dependence of the coupling constants. This universal dependence leaves no means of tuning the rate through the phonon environment. 

The phonon-induced spin-flip rate is jointly dependent on the local phonon density of states (DoS) and on the symmetry matching between the mode profile and transition selection rules. In devices built on structured mechanical environments, the spin lifetime that limits the device functionality therefore depends on the particular structure of the phonon environment and cannot be inferred from bulk studies, opening up pathways to control the relaxation process. 

Already the simplest modification of the DoS via geometric confinement of phonons alters the spin relaxation dynamics. Spin relaxation induced by confined acoustic modes has been analyzed for a QD in a suspended slab \cite{Liao_2006} and for nanowire-based QDs \cite{Yin_2010}, where van Hove singularities of the confined branches already produce sharp features in the rate. More elaborate architectures, such as phononic crystals, further enhance the flexibility of control over the DoS \cite{lutz_2016,Odeh_2025}. Engineered phononic environments have been explored to suppress the orbital relaxation of defect centers \cite{Klotz_2022,Kuruma_2025} and proposed to inhibit spin flips of rare-earth ensembles \cite{lutz_2016}, but the structured-phonon literature has otherwise emphasized phononic Purcell factors and the direct mechanical driving of the spin \cite{Joe2026}. Experiments already hint that the gap, rather than confinement alone, is the operative ingredient: in etched diamond nanopillars, which reduce the phonon density of states without opening a gap at the transition frequency, the predicted lifetime enhancement was not observed, being masked by fabrication damage and by leakage into the substrate continuum \cite{Lutz_2024}. To our knowledge, however, for the particular system of an electron spin confined in a QD embedded in a PnC, the phonon-induced spin flip has been studied neither theoretically nor experimentally. Whether an elastic band gap suppresses the spin-flip rate, whether van Hove peaks enhance it, and how mode symmetry selects which strain components couple to the spin remain open questions.

In this work, we derive a general theory of phonon-induced relaxation of a single electron in an InGaAs QD embedded inside a phononic crystal waveguide. Our goal is to present a comprehensive discussion not only of the DoS and the resulting Fermi-golden-rule relaxation rates, but also of the interplay between mode symmetries, field orientations, QD positions, and selection rules, in particular near the van Hove singularities, where non-Markovian dynamics emerges. 

The elastic eigenmodes are obtained from a finite-element solution and classified according to symmetry. We resolve the Markovian spin-flip rate by mode symmetry, magnetic-field orientation, and the position of the QD within the waveguide core. We show that the spin relaxation rate mostly exceeds the bulk value but is suppressed by many orders of magnitude in certain ranges of magnetic field. At the Brillouin zone (BZ) center and edges, in contrast, slow-mode effects lead to singularities in the relaxation rate. Here a power-law analysis separates the universal density of states divergence from the mode-specific behavior of the strain coupling. The resulting branch- and orientation-dependent exponents suppress some divergences and reinforce others; in the latter case, spin relaxation is substantially enhanced in a narrow range of fields around the singularity. 
Near these van Hove singularities the Born--Markov description breaks down: the bath memory time and the spectral density both diverge as the group velocity vanishes, so a spin tuned to a singularity needs to be treated beyond the Markov approximation. 
For this, we follow the resolvent approach \cite{WoldeyohannesJohn2003, YangJohn2007, Roy2010, lutz_2016} based on the Weisskopf--Wigner theory~\cite{Weisskopf_1930}, generalized to finite temperature. We show that in the direct vicinity of a singularity, exponential relaxation is replaced by a universal power-law decay to a finite constant occupation in which the spin is dressed into the slow mode to form a ``spin polaron''. 

The dynamical behavior revealed by our analysis, unparalleled in bulk systems, enables the modulation of spin decay rates in phononic structures over several orders of magnitude through tunable external parameters. This paves the way for new spin control schemes, including hybrid protocols involving mechanical degrees of freedom.
In addition, our results demonstrate large spatial inhomogeneity of relaxation properties and the emergence of non-Markovian dynamics near DoS singularities.
Although we have chosen to study a QD system, our theory is applicable to any spin or orbital few-level system resonant with guided phonon modes, as long as the coupling Hamiltonian is known either on a microscopic or effective level.

The paper is organized as follows. Section~\ref{sec:system} defines the geometry and material parameters of the waveguide. Section~\ref{sec:theory} sets up the effective spin-phonon Hamiltonian, reduces it, within a Born--Markov--secular Lindblad treatment, to a strain-resolved expression for the spin-flip rate, and develops a finite-temperature Weisskopf--Wigner treatment valid at the van Hove singularities, where the Markov approximation fails. Section~\ref{sec:results} presents the results for dispersion, the strain profiles, rates resolved by mode class and field orientation, the power-law analysis at the BZ extrema, and the non-Markovian dynamics at finite detuning from the singularities. Section~\ref{sec:conclusions} summarizes the findings. Some technical aspects are further analyzed in the Appendices.

\section{System and model}\label{sec:system}
%

While the theoretical framework to be developed below is general, we choose a W1m line-defect waveguide in a free-standing GaAs slab as a concrete setting. Its guided modes carry definite parity under the in-plane and out-of-plane mirror operations of the slab, and the resulting strain profiles split into four symmetry classes that couple to the spin selectively under different magnetic-field orientations.

\begin{figure}[tb]
  \centering
  \makebox[0.75\linewidth][l]{(a)}%
  \makebox[0.20\linewidth][l]{(b)}\\
  \includegraphics[width=2.7095in]{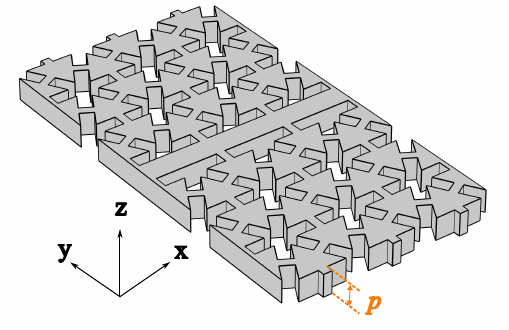}%
  \hspace{3pt}%
  \includegraphics[width=0.6240in]{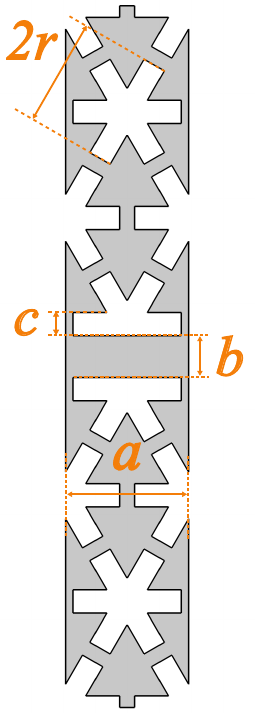}%
  
    \caption{Schematic of W1m GaAs phononic crystal. (a) Isometric view of three supercells of the W1m snowflake phononic crystal waveguide fabricated from a $(001)$-oriented GaAs slab of thickness $p = 220$ nm. (b) Top view of a supercell with lattice constant $a= 760 \, \mr{nm}$, snowflake radius $r= 0.44a$, arm width $c = 0.19a$, and the channel width $b = \sqrt{3} \,a - 2\Delta -c$, where $\Delta = 0.6 a$ is the inward shift of the adjacent rows.}
    \label{fig:waveguide_schematics}
\end{figure}

We consider a single electron spin confined in a self-assembled InGaAs QD embedded in a phononic crystal waveguide fabricated from a (001)-oriented GaAs slab. We adopt the crystallographic coordinate system $x||[110]$, $y||[\overline{1}10]$, $z||[001]$, consistent with \cite{rosinski_2026}. The PnC has a snowflake lattice geometry with lattice constant $a= 760 \, \mr{nm}$, snowflake radius $r= 0.44a$, arm width $c = 0.19a$, and slab thickness $p = 220 \, \mr{nm}$ in the $z$-direction, as labeled in ~\subfigsref{fig:waveguide_schematics}{a}{b}. The waveguide is of W1m type ~\cite{rosinski_2026} formed along $x$ by removing a row of snowflake holes, as shown in \subfigref{fig:waveguide_schematics}{b}. The two adjacent snowflake rows are shifted inward by $\Delta = 0.6a$, so the channel width is $b=\sqrt{3} \,a - 2\Delta -c$. This setup produces a phononic band gap in the frequency range $2.15-3.33\, \mr{GHz}$, within which multiple guided modes are confined to the waveguide region \cite{rosinski_2026}.

The QD is modeled as a point-like object, located at position $\rr_0$ within the waveguide, with $\rr_0=(0,0,0)$ corresponding to the channel center at the slab mid-plane.  

\par
%
%
The phononic modes of the waveguide are computed numerically in the frequency domain by finite-element simulations ~\cite{comsol}. The underlying model solves the coupled equations of linear elasticity and piezoelectricity ~\cite{rosinski_2026}. The supercell consists of a single unit cell $a$ in the propagation direction ($x$) and four rows of the snowflake lattice on each side of the channel in $y$-direction, where the row adjacent to the channel has had its two arms extending into the core removed (the W1m modification), followed by two complete snowflake rows, and a partial row at the boundary. 

Floquet boundary conditions are applied in the propagation direction $x$ with Bloch wave vector $\vk$, periodic boundary conditions in the $y$-direction, and traction-free surfaces on the remaining surfaces. The mesh is partitioned with mirror symmetry about the $y=0$ plane to ensure correct symmetry classification of near-degenerate modes.
Each guided mode is classified by a two-letter symmetry label (SS, SA, AS, AA) denoting the symmetry (S) and antisymmetry (A) of its displacement field under $z$ and $y$-mirror reflections, respectively. Guided modes are identified by their energy confinement to the waveguide channel.
The spin-phonon coupling is described by the effective spin-phonon Hamiltonian 
\cite{vanvleckParamagneticRelaxationTimes1940,rothFactorDonorSpinLattice1960}
\begin{equation}\label{eq:H_sph}
H_{\mr{sph}}(\rr) = \frac{1}{2} \mu_{\mr{B}} B_i \delta \hat{g}_{ij} \sigma_j,
\end{equation}
which accounts for the leading spin-relaxation mechanism in self-assembled QDs \cite{mielnik-pyszczorski_2018}. 
Here $\mu_{\mr{B}}$ is the Bohr magneton, $B_i$ are the components of the applied magnetic field, $\sigma_j$ are Pauli matrices, and the strain-induced g-factor modification is $\delta\hat{g}_{ij} = \eta\eps_{ij}(\rr)$, with
\begin{equation}\label{eq:eta}
    \eta = \left[ \frac{d_{\mr{v}}}{\sqrt{3} \, E_{\mr{g}}} -\frac{1}{3} \right] \frac{2 \, E_{\mr{P}} \, \Delta_{\mr{so}}}{E_{\mr{g}}(E_{\mr{g}} + \Delta_{\mr{so}})}.
\end{equation}

The modification $\delta \hat{g}_{ij}$ contains two contributions, referred to as off-diagonal and deformation-potential after the distinct ways in which they emerge from the multiband $\kp$ Hamiltonian  ~\cite{mielnik-pyszczorski_2018}. $\Delta_{\mr{so}}$ is the spin--orbit splitting of the valence band, $E_{\mr{g}}$ is the bulk band gap, and $d_{\mr{v}}$ is the valence-band deformation potential.  Here $E_{\mr{P}}$ is the Kane energy parameter ~\cite{Vurgaftman_2001}, 
\begin{equation}
    E_{\mr{P}}=\left( \frac{m_{\mr{0}}}{m_{\mr{e}}^*}-1 \right) \frac{E_{\mr{g}}(E_{\mr{g}}+\Delta_{\mr{so}})}{E_{\mr{g}}+(2/3)\Delta_{\mr{so}}}.
\end{equation}

\section{Theory of spin-phonon coupling for the waveguide modes}
\label{sec:theory}

This section develops the theoretical framework connecting the quantized phonon modes of the waveguide to the electron spin. We first quantize the guided elastic field and project the spin-phonon Hamiltonian [\eqnref{eq:H_sph}] onto the confined QD ground state to obtain the mode-resolved spin-flip coupling constant. We then derive the Markovian spin-flip rate from this coupling within the Born--Markov--secular approximation, expressing it through a spectral density resolved by branch, field orientation, and QD position. Finally, we extend the treatment beyond the Markov approximation to the non-Markovian dynamics that emerge when the Zeeman frequency approaches a van Hove singularity of the waveguide dispersion.
  
%
\subsection{Eigenmode expansion}
\label{sec:theory-formalism}
The eigenmodes of the waveguide satisfy the eigenvalue equation
\begin{equation}\label{eq:eigenvalue}
    -\rho(\rr)\omega^2_{\lk} u^{\lk}_i(\rr)= \partial_j C_{ijmn} \partial_n u^{\lk}_m(\rr),
\end{equation}
where $\lambda$ is the mode index, $k$ denotes the Bloch wave vector along the propagation direction, $C_{ijmn}$ is the elastic stiffness tensor, and $\rho(\rr)$ is the crystal  density. Einstein's summation convention over repeated indices is used throughout the paper. Here $u^{\lk}_i(\rr)$ denotes the Cartesian components of the displacement eigenmode $\bm{u}^{\lk}(\rr)$. \aeqnref{eq:eigenvalue} has the structure of a Hermitian eigenvalue problem $\hat{D} \bm{u} = \omega^2 \bm{u}$ with respect to the inner product
\begin{equation}\label{eq:inner_product}
\langle\bm{u}|\bm{v}\rangle = \frac{1}{M}\int_V d^3r\, \rho(\rr)\, u_i^*(\rr)\, v_i(\rr),
\end{equation}
where $M$ is an arbitrary mass introduced to render the eigenmodes dimensionless, which can be chosen as 
$M=\int_V d^3r\, \rho(\rr)$,
and $V$ is the volume of the supercell. 
For a one-dimensional waveguide, the eigenmodes satisfy the Bloch condition $\bm{u}^{\lk}(\rr + a \hat{x})=e^{ika} \bm{u}^{\lk}(\rr)$, where $\hat{x}$ is the unit vector along the propagation direction, and the relation $[\bm{u}^{\lk}(\rr)]^* = \bm{u}^{\lnk}(\rr)$ holds since the eigenvalue equation has real coefficients. 

The spin-phonon Hamiltonian in Eq.~\eqref{eq:H_sph} is projected onto the electron orbital ground state and expanded in the quantized eigenmodes of the phononic structure with the annihilation (creation) operators $b_{\lck}$ ($b_{\lcnk}^{\dagger}$) (see Appendix~\ref{sec:app-quantize} for details). Here it is convenient to choose the $z$-axis along the magnetic field, so that $B_z = B$ is the only non-vanishing component. Then, written in terms of raising and lowering operators $\sigma_{\pm}=(\sigma_x \pm i\sigma_y)/2$, the Hamiltonian takes the form
\begin{equation}\label{eq:H_sph_alpha}
H_{\mr{sph}} = \sum_{\lck} \alpha(\lck)\, \sigma_+ \, (b_{\lck} + b_{\lcnk}^\dagger) \, + \, \hc,
\end{equation}
where 
\begin{equation}\label{eq:alpha_Bz}
\alpha(\lck) =\frac{1}{2}  \sqrt{\frac{\hbar}{2 M \omega_{\lk}}} \eta   \mu_{\mr{B}} B  \Bigl [ \eps_{xz}^{\lk}(\rr_0) -i\eps_{yz}^{\lk}(\rr_0) \Bigr],
\end{equation}
with 
\begin{equation}\label{eq:eps-symmetry}
\eps_{ij}^{\lk}(\rr)= \left[\eps_{ij}^{\lambda,-k}(\rr)\right]^* =
\frac{1}{2}\left[\partial_i u^{\lk}_j(\rr)+\partial_j u^{\lk}_i(\rr)\right]    
\end{equation}
denoting the strain pertaining to the eigenmode $\lk$. 

\subsection{Spectral density and Markovian transition rates}
\label{sec:Spectral density and transition rates}

Under standard conditions (to be discussed in detail in the following), 
the spin dynamics are governed by the Lindblad master equation ~\cite{HBreuer&FPetruccione}. Within the Born--Markov and secular approximations, the rate of the transition from the upper to the lower Zeeman sublevel is
\begin{equation}
\gamma_\doa = 2 \pi  \, \left[n_{\mr{B}} \left(\frac{\Delta E_{\mr{Z}}}{\hbar}\right)+1 \right] \, \zeta\left(\frac{\Delta E_{\mr{Z}}}{\hbar}\right), \label{eq:gamma_down}
\end{equation}
where $\Delta E_{\mr{Z}} = |g| \mu_{\mr{B}} B > 0$ is the Zeeman energy splitting, $n_{\mr{B}}(\omega)=[\exp(\hbar\omega / k_{\mr{B}} T)-1]^{-1}$ is the Bose occupation number, and $\zeta(\omega)$ is the spectral density of the spin-phonon coupling, defined below. We denote the Zeeman angular frequency by $\omz = \Delta E_{\mr{Z}}/\hbar$ and the corresponding ordinary frequency by $f_Z = \omz/2\pi$.

For the spin-phonon Hamiltonian of \eqnref{eq:H_sph}, the spectral density takes the form
\begin{equation}\label{eq:zeta_1D}
\zeta(\omega)  = \frac{1}{\hbar^2} \sum_{\lck} |\alpha(\lck)|^2  \delta(\omega - \omega_{\lk}).
\end{equation}
Converting the sum over $k$ to an integral, $\sum_k \rightarrow (L/2\pi) \int dk$, where $L$ is the quantization length of the waveguide along the propagation direction, and performing the integration over the delta function yields
\begin{align}\label{eq:zeta_final}
\zeta(\omega) &= \frac{1}{8 \pi} \frac{ \mu_{\mr{B}}^2 }{\hbar } \frac{L}{M}
  \eta^2 B^2 \frac{1}{\omega} \notag \\ \times &\sum_{ \lambda} \Big( \Bigl | \eps_{xz}^{\lambda,k_{\lambda \omega }}(\rr_0) \Bigr|^2 + \Bigl|\eps_{yz}^{\lambda,k_{\lambda \omega }} (\rr_0) \Bigr|^2  \Big) \left|\frac{dk}{d\omega}\right|_{{k_{\lambda \omega }}},
\end{align}
where $k_{\lambda \omega}>0 $ runs over all wavevectors satisfying $\omega_{\lk}=\omega$ on branch $\lambda$. The sum accounts for all the branches that intersect a given frequency, for which the solution $\omega_{\lk}=\omega$ exists. The solutions come in pairs $\pm k_{\lambda \omega}$, and here both have already been included using the symmetry from Eq.~\eqref{eq:eps-symmetry}. The factor $|dk/d\omega|$ is the one-dimensional density of states per branch at that point.  

We consider the Faraday configuration ($\vB \parallel \hat{z}$) and two Voigt configurations ($\vB \parallel \hat{x}$ and $\vB \parallel \hat{y}$), where we again refer to the sample coordinate plane defined in Sec.~\ref{sec:system}. To align this reference frame with that used in the formalism of Sec.~\ref{sec:theory-formalism}, the Voigt configurations are obtained by applying a corresponding rotation to the spin operators. In the spin-phonon Hamiltonian [\eqnref{eq:H_sph_alpha}], the spin-flip transitions then couple to $|\eps_{xz}|$ and $|\eps_{yz}|$ for $\vB \parallel \hat{z}$, to $|\eps_{xy}|$ and $|\eps_{xz}|$ for $\vB \parallel \hat{x}$, and to $|\eps_{yz}|$ and $|\eps_{xy}|$ for $\vB \parallel \hat{y}$. 

The Lindblad master equation follows from a separation of timescales between the phonon bath and the spin dynamics. The phonon bath correlation function decays on a memory time $\tau_{\mr{mem}} \sim \ell / v_{\mr{g}}$~\cite{muljarov2004}, where $\ell$ is the QD size and $v_{\mr{g}}= d\omega/dk$. The secular approximation requires $\omz^{-1} \ll \gamma_{\doa}^{-1}$, where $\gamma_{\doa}^{-1}$ is the spin relaxation time, and the Born--Markov approximation requires $\tau_{\mr{mem}} \ll \gamma_{\doa}^{-1}$. In bulk GaAs, with $\ell \approx 4$ nm and speed of sound between $2.85 \times 10^{3}$ and $5.24 \times 10^{3}$ m/s, one finds $\tau_{\mr{mem}} \approx 1$~ps, which is well below $\gamma_{\doa}^{-1}$. In the waveguide, $v_{\mr{g}}$ typically ranges from $\sim 10^2$ to $4 \times 10^{3}$ m/s across the guided branches within the band gap, as will be seen in Sec.~\ref{sec:results}. Therefore, $\tau_{\mr{mem}}$ grows to a few picoseconds in the waveguide, and to tens of picoseconds in the slow-velocity tail near the BZ center and edges, but remains much shorter than $\gamma_{\doa}^{-1}$ across most of the band gap. The Born--Markov condition thus holds across most of the band gap, degrading only as the BZ center and edges are approached, where $v_{\mr{g}} \to 0$. At those points, the spectral density diverges as a consequence of the vanishing $v_{\mr{g}}$. The secular condition is satisfied away from these points. Away from the BZ center and edges, the rates remain well below $0.1\,\mr{s}^{-1}$ across all QD positions and field orientations considered, so $\gamma_{\doa}^{-1} \gg \omz^{-1}$ holds within the regime where the Born-Markov approximation is valid, as shown in Sec.~\ref{sec:faraday} and Sec.~\ref{sec:voigt}. 

\subsection{Non-Markovian dynamics at the van Hove singularities}
\label{sec:ww}

As the BZ center and edges are approached, the memory time of the phonon bath diverges, and the Born--Markov treatment of Section~\ref{sec:Spectral density and transition rates} fails. We now solve the spin dynamics at or near the singularity without the Markov approximation, at finite temperature. Among the available non-Markovian techniques \cite{deVega2017}, we follow the resolvent (Weisskopf--Wigner) approach previously applied to optical transitions near a photonic band edge \cite{WoldeyohannesJohn2003, YangJohn2007, Roy2010}.

We drop the counter-rotating terms $\sigma_+ b^\dagger_{\lcnk}$ and $\sigma_- b_{\lcnk}$ from the spin-phonon Hamiltonian derived above, so that the total Hamiltonian reads
\begin{align}
H ={}& \frac{\hbar\omz}{2}\sigma_z + \sum_{\lck}\hbar\omega_{\lk}\, b^\dagger_{\lck} b_{\lck} \notag\\
&+ \sum_{\lck}\bigl[\alpha(\lck)\,\sigma_+ b_{\lck} + \alpha^*(\lck)\,\sigma_- b^\dagger_{\lck}\bigr].
\label{eq:ww_H}
\end{align}
We initialize the bath in a Fock state with fixed occupations $\{n_{\lk}\}$ and the spin in the upper Zeeman level. The coupling in \eqnref{eq:ww_H} conserves the total excitation number, so the state is restricted to
\begin{align}
\lvert\Psi(t)\rangle ={}& \tilde C_e(t)\,\lvert\upa;\{n_{\lk}\}\rangle \notag\\
&+ \sum_{\lck} \tilde C_{\lk}(t)\,\lvert\doa;\{n_{\lk}\}+1_{\lk}\rangle,
\label{eq:ww_ansatz}
\end{align}
where the tildes denote interaction-picture amplitudes and $\tilde C_e(0)=1$, $\tilde C_{\lk}(0)=0$. 
In the following we summarize the solution for the survival amplitude of the initial state $C_e(t)$ and discuss the resulting dynamics, while Appendix~\ref{sec:app-non-markov} provides the details of the derivation and links the dynamics to the analytical structure of the solution.

Substituting \eqnref{eq:ww_ansatz} into the Schr\"odinger equation with the Hamiltonian given by \eqnref{eq:ww_H} yields a coupled set of equations of motion. Following the standard approach, the equation for $\tilde{C}_{\lk}(t)$ is formally integrated and substituted into the other one, to produce a Volterra integro-differential equation. This is solved by Laplace transform in the parabolic approximation near a band extremum (van Hove singularity), $\omega_{\lk} = \omega_0 + \beta(k-k_0)^2$. For $\beta > 0$ (dispersion minimum, for definiteness) and $\omz \ge \omega_0$ (transition energy within the band) this yields the Laplace transform $\tilde C_e(s)$ of the amplitude
\begin{equation}
\tilde C_e(s) = \frac{1}{s - i\pi\mathcal{A}/\sqrt{-\Delta_0 - is}}, \quad
\mr{Re}\sqrt{-\Delta_0 - is} \ge 0,
\label{eq:ww_resolvent}
\end{equation}
with the detuning of the Zeeman frequency from the van Hove singularity $\Delta_0 = \omz - \omega_0$, and
\begin{align}\label{eq:A-prefactor}
\mathcal{A} ={}& \bigl[n_{\mr{B}}(\omega_0)+1\bigr]\,
\frac{\mu_{\mr{B}}^2\,\eta^2 B_z^2\,L}{16\pi\,\hbar\,M\,\omega_0\,\beta^{1/2}} 
\notag \\
&\times\sum_{\lambda}\Bigl(\bigl|\eps_{xz}^{\lambda k_0}(\rr_0)\bigr|^2
+ \bigl|\eps_{yz}^{\lambda k_0}(\rr_0)\bigr|^2\Bigr),
\end{align}
written in the coordinate frame aligned to magnetic field as defined in Sec.~\ref{sec:theory-formalism}.

The time-domain dynamics of the upper-state amplitude $\tilde{C}_{e}(t)$ is recovered by the inverse Laplace transform (see Appendix~\ref{sec:app-non-markov} for details of the derivation). This is done in the transformed variable $z=-\Delta_0-is$. Setting $w=z^{1/2}$, the poles of $\tilde C_e$ are found from the cubic equation $w^3 + \Delta_0 w - \pi\mathcal{A} = 0$, whose single positive real root $w_0$ corresponds to an oscillating (undamped) component of the solution and gives the pole position $z_0 = w_0^2$ and hence the bound-state frequency $\Omega = \Delta_0 + z_0 = \Delta_0 + w_0^2$. A complex-conjugate pair of roots $w_{1,2} = -w_0/2 \pm i\chi$ defines the poles at $z_{1,2} = w_{1,2}^2$ that lie on the second Riemann sheet in $z$. The integration contour can be deformed to capture the pole at $z_2$ (with $\mr{Im}\,z_2 > 0$) \cite{sudarshanDecayingStates1978,nakazato96}, which yields a transient contribution to the occupation $|\tilde C_e|^2$ decaying exponentially at the rate $\Gamma' = 2\,\mr{Im}\,z_2 = 2w_0\chi$, where $\chi = \sqrt{\Delta_0 + \tfrac{3}{4}w_0^2}$. The remaining part of the deformed contour yields a power-law tail (see Appendix~\ref{sec:app-non-markov} for details). Thus, the survival amplitude decomposes as
\begin{equation}
\label{eq:ww_decomposition}
\tilde{C}_e(t) = Ze^{i \Omega t} + Z' e^{i(\Delta_0 + z_2) t} 
+ I_{\text{rem}}(t), 
\end{equation}
where
\begin{equation}\label{eq:ww_Z}
Z=\frac{2w_0^2}{3w_0^2+\Delta_0}
\end{equation}
is the bound-state weight, $Z' = 2w_2^2/(3w_2^2+\Delta_0)$ the transient weight, and $I_{\text{rem}}(t)$ is a non-exponential remainder [see Eq.~\eqref{eq:I-rem} in Appendix~\ref{sec:app-non-markov}]. 

Normalization requires $Z + Z' + I_{\text{rem}}(0) = 1$.
As $\Delta_0$ becomes large, we find $Z' = 1 + \mathcal{O}(\pi\mathcal{A}/\Delta_0^{3/2})$, so that the solution is dominated by the exponential decay, until the power-law asymptotics sets in at long times (hence we refer to it as \textit{transient}). 
In the long-time regime, $w_0^2 t \gg 1$, the exponentially damped term is negligible, while the remainder $I_{\text{rem}}(t)$ decays as $t^{-3/2}$ [Eq.~\eqref{eq:I-rem-longt} in Appendix~\ref{sec:app-non-markov}], so that the long-time survival probability of the upper Zeeman level follows as
\begin{align}
|\tilde{C}_e(t\to\infty)|^2 \simeq{}& \frac{4w_0^4}{(3w_0^2 + \Delta_0)^2} + \frac{1}{4\pi^3 \mathcal{A}^2 t^3} \notag\\
&+ \frac{2w_0^2}{\pi^{3/2} \mathcal{A}\, t^{3/2}(3w_0^2 + \Delta_0)} \notag\\
&\times\cos\!\left[(\Omega - \Delta_0)\,t + \frac{3\pi}{4}\right].
\label{eq:ww_result}
\end{align}
A population fraction $|Z|^2$ remains permanently locked in the polaron-like bound state. The second term is the released occupation decaying via the algebraic tail, and the third is an interference term that modulates the approach to the asymptotic plateau.

Eq.~\eqref{eq:ww_result} explicitly shows that the long-time dynamics shows no asymptotic exponential decay at any detuning. This is to some extent similar to the electron-LO-phonon coupling of QD electrons, where the flat dispersion of LO phonons at the zone center binds a collective phonon mode to the charge carrier, forming a long-lived polaron, which precludes exponential relaxation \cite{Hameau_1999,Guldner_2001}. The exponential behavior emerges, however, as a transient described by the $Z'$ term in Eq.~\eqref{eq:ww_decomposition}, extending to longer times as the detuning $\Delta_0$ grows. 
Its rate $\Gamma'$ approaches the Markovian golden-rule value as $\Delta_0/D_0^{*}\to\infty$ (Sec.~\ref{sec:app-gr_convergence} of Appendix~\ref{sec:app-non-markov}). 

As we discuss in Appendix~\ref{sec:app-non-markov}, the system dynamics allows scaling with 
the parameter
\begin{equation}
D_0^{*} = (\pi\mathcal{A})^{2/3},
\label{eq:ww_D0}
\end{equation}
which sets the van Hove frequency scale.
All dimensionless quantities: $\Omega/D_0^{*}$, $Z$, $\Gamma'/D_0^{*}$, and the dynamics as a function of $D_0^{*}t$ depend on the detuning only through $\delta = \Delta_0/D_0^{*}$. At $\delta = 0$ one finds $Z = 2/3$, $Z' + I_{\text{rem}}(0) = 1/3$, and a trapped population $|Z|^2 = 4/9$, while for $\delta \gg 1$ the bound state carries the asymptotically small weight $Z \simeq 2(D_0^{*}/\Delta_0)^{3}$.
Temperature enters the dynamics exclusively through the scaling parameter $\mathcal{A} \propto n_{\mr{B}}(\omega_0)+1$. An increase in temperature increases $D_0^{*}$, moves the bound-state pole at $z_0 = w_0^2$, which lies outside the band below the band-edge frequency $\omega_0$ at every detuning, further from the band edge, widens the non-Markovian window, and alters the trapped population $|Z|^2$, but it does not restore exponential relaxation. 

The opposite case of negative detuning, in which the Zeeman frequency lies inside the band gap, is analyzed in Sec.~\ref{sec:app-negdet} of Appendix~\ref{sec:app-non-markov}. 

\section{Results}\label{sec:results}

This section presents the numerical results for the phonon-induced spin-flip rate of an InGaAs quantum dot in the W1m waveguide. We first characterize the guided-mode dispersion and the strain profiles that set the spin-phonon coupling for each symmetry branch. We then evaluate the Markovian spin-flip rate in the Faraday and Voigt configurations, analyze its power-law behavior at the van Hove singularities, and finally treat the non-Markovian dynamics that arise at finite detuning from these singularities.
For numerical evaluation we set $g=-2$. The rates scale as $B^2$ at fixed Zeeman frequency, so results for any other $g$-factor follow by multiplying by $(2/|g|)^2$, and in experiment $\Delta E_{\mr{Z}}$ is measured directly. All rates are evaluated at $T = 4\,\mr{K}$.

In our simulations we use the material parameters listed in \tabref{tab:parameters}. Note that the QD band parameters and deformation potential enter only via the parameter $\eta$ in Eq.~\eqref{eq:eta}. Since these parameters are not precisely defined in a strained, alloyed, and confined system, we simply take the GaAs values as an estimate.

\begin{table}[tb]
    \centering
    \begin{tabular}{l|c|c}
        Band gap energy & $E_{\mr{g}}$ & $1.519 \, \mr{eV}$\\
        Effective electron mass & $m_{\mr{e}}^*$ & $0.0665 \, m_{\mr{0}}$\\
        spin--orbit splitting & $\Delta_{\mr{so}}$ & $0.341 \, \mr{eV}$ \\
        Deformation potential& $d_{\mr{v}}$ & $-4.8 \, \mr{eV}$ \\
        \multirow{3}{*}{Elastic stiffness} & $C_{11}$ & $122.1 \,  \mr{GPa}$\\
        & $C_{12}$ & $56.6 \,$GPa\\
        & $C_{44}$ & $60 \, $GPa\\
        Piezoelectric constant ~\cite{Caro_2015} & $e_{14}$ & $-0.205 \, \mr{C \cdot m^{-2}}$\\
        Dielectric constant ~\cite{Winkler_2003} & $\kappa$  & $12.4$\\
        Crystal density ~\cite{Brice_1996} & $\rho$  & $5332 \, \mr{kg \cdot m^{-3}}$   \\ 
    \end{tabular}
    \caption{Material parameters for GaAs used in this work. Unless stated otherwise, the values are obtained from \cite{Vurgaftman_2001}.}
    \label{tab:parameters}
\end{table}

\subsection{Waveguide dispersion and mode structure}\label{sec:dispersion}

\begin{figure}[tb]
    \centering
    \includegraphics[width=1.0\linewidth]{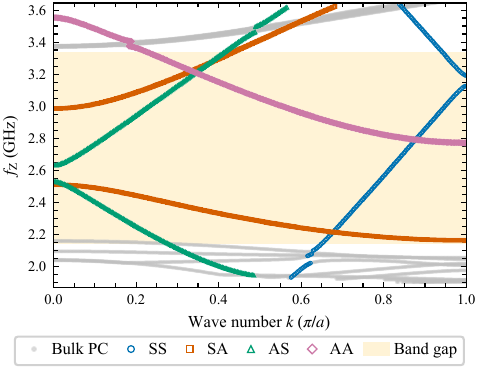}
    \caption{Phonon dispersion of the W1m waveguide. Guided modes within the phononic band gap (shaded) are labelled as SS, SA, AS, and AA. Gray curves show the bulk modes.}
    \label{fig:dispersion}
\end{figure}

\afigref{fig:dispersion} shows the dispersion of the guided phonon
branches inside the band gap. Each branch is labeled by its symmetry
class (SS, SA, AS, or AA), and this classification fixes which strain components the mode carries and, therefore, which field orientations couple it to the spin (Section~\ref{sec:Strain_profiles}). At the BZ center and at the BZ edges, the group velocity $v_{\mr{g}}$ vanishes, producing van Hove singularities in the DoS, which scales as $v_{\mr{g}}^{-1}$. These singularities drive the enhancements at the BZ center and edges, analyzed in
Section~\ref{sec:powerlaw}.

A narrow interval between $\sim 2.53$ and $\sim 2.63$~GHz contains only SS branches. These modes carry only the diagonal strains $\eps_{xx}$, $\eps_{yy}$, and $\eps_{zz}$, which do not couple to the spin in either the Faraday or the Voigt configuration. The lattice vibrations in this interval therefore generate no spin-coupled strain, and the spin remains effectively isolated from the phonon environment at every QD position considered. Below we refer to this interval as the SS-only protected window. Coupling through diagonal strains could, in principle, be activated under tilted field orientations via their non-volumetric (axial) combination $2\eps_{zz}-\eps_{xx}-\eps_{yy}$, which we find to be non-vanishing at the QD positions, but we do not address that case here.

\subsection{Strain profiles and mode-field coupling}\label{sec:Strain_profiles}

\begin{figure}[th]
    \centering
    \includegraphics[width=1.0\linewidth]{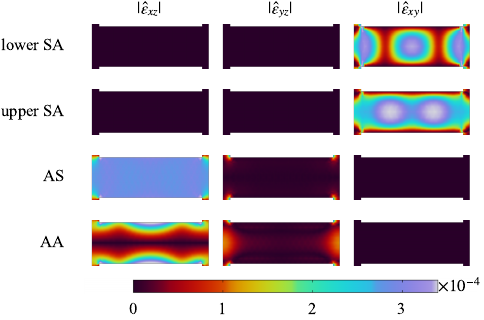}
    \caption{Normalized shear strain amplitude $|\eps_{xz}|$, $|\eps_{yz}|$, and $|\eps_{xy}|$ at mid-plane $z=0$ and $k=0.37209(\pi/a)$. Rows show the lower SA, upper SA, AS, and AA  modes $(2.3483,\,3.2616,\,3.2595,\,3.1778)$ GHz, respectively. Only the waveguide core area across one unit cell is shown. The color bar is shared across all panels. Strain amplitudes are normalized to a maximum displacement of $0.1$ nm.}
    \label{fig:strain_maps_z0}
\end{figure}

\begin{figure}[th]
    \centering
    \includegraphics[width=1.0\linewidth]{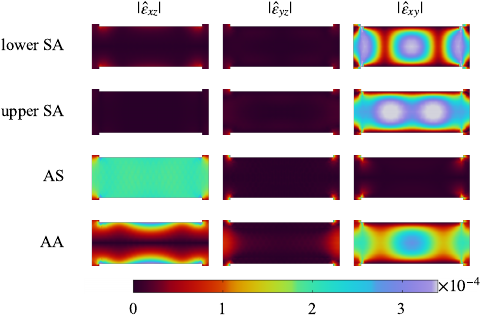}
    \caption{Same as \figref{fig:strain_maps_z0} at $z=z_{\max}=60$ nm, the maximum QD displacement from the mid-plane consistent with a minimum $50$ nm distance from the slab surface. }
    \label{fig:strain_maps_zmax}
\end{figure}

The shear strain components determine the spin-phonon coupling for the SA, AS, and AA branches. \figref{fig:strain_maps_z0} and \figref{fig:strain_maps_zmax} display the shear strain amplitudes $|\eps_{xz}|$, $|\eps_{yz}|$, and $|\eps_{xy}|$ of the mass-normalized eigenmodes at a representative mid-gap wave vector, at the mid-plane $z=0$ and at $z=z_{\max}=60$~nm, respectively. The displacement $z_{\max}$ is the largest vertical offset of the QD from the mid-plane consistent with a minimum distance of $50$~nm from the slab surface.

The SS branches are omitted. Their spin-flip-active strain components vanish by symmetry at the QD positions and field orientations considered (Section~\ref{sec:dispersion}), so their contribution to the spin-flip rate is exactly zero. The numerically computed values remain below $10^{-6}\,\mr{s}^{-1}$ across the full band gap, a residual attributable to the finite discretization of the simulation.

The SA mode is symmetric under $z \to -z$, which forces $\eps_{xz}$ and $\eps_{yz}$ to vanish at the mid-plane and leaves $|\eps_{xy}|$ as the dominant component [\figref{fig:strain_maps_z0}, top two rows]. The two SA branches differ in their parity under reflection $x \to -x$, which sets their nodal structure across the waveguide core.

The AS mode is antisymmetric under $z \to -z$, which places the $\eps_{xz}$ antinode at the mid-plane and forces $\eps_{xy}$ to vanish there [\figref{fig:strain_maps_z0}, third row]. Its additional symmetry under $y \to -y$ suppresses $u_y$ near the core, so $|\eps_{yz}|$ is also negligible. The resulting $|\eps_{xz}|$ profile fills the core uniformly and is preserved at $z_{\max}$ [\figref{fig:strain_maps_zmax}, third row]. Vertical displacement of the QD, therefore, only rescales the AS coupling, leaving its insensitivity to in-plane position intact.

The AA mode carries $|\eps_{xz}|$ at the mid-plane, with a node along
the centerline $y=0$ [\figref{fig:strain_maps_z0}, bottom row] and antinodes near the lateral edges of the core. This nodal line follows from the $y$-antisymmetry of the mode:
$u_x$ and $u_z$ are both odd in $y$, so $\eps_{xz}$ is also odd in $y$ and vanishes at
$y=0$. The complementary component $\eps_{xy}$ is forced to vanish at
$z=0$ by the $z$-antisymmetry, since $u_x$ and $u_y$ are both odd in
$z$. As a result, $|\eps_{xy}|$ is negligible at the mid-plane and
develops appreciable amplitude only at $z_{\max}$, where it acquires its own nodal structure across the channel while $|\eps_{xz}|$ recedes toward the lateral edges. The AA coupling is therefore suppressed at the channel center and activated by displacing the dot.

\par

\subsection{Spin-flip rates in the Faraday configuration}\label{sec:faraday}

\asubfigsref{fig:bz_by_comparison}{a}{b} show the spin-flip rate $\gamma_{\doa}$ versus the Zeeman frequency $f_Z$ in the Faraday configuration, at the channel center and at the lateral displacement $(0,b/4,0)$.

At the channel center [\asubfigref{fig:bz_by_comparison}{a}], the AS
mode dominates $\gamma_{\doa}$ across the entire band gap, except
inside the SS-only protected window of Section~\ref{sec:dispersion}, where the SS branches are the sole contributor. The AA contribution remains weak throughout. This hierarchy reflects the strain profiles of \figref{fig:strain_maps_z0}. Sharp peaks at the BZ center and edges arise from van Hove singularities in the DoS and are analyzed in Section~\ref{sec:powerlaw}.

Displacing the QD to $(0,b/4,0)$ [\asubfigref{fig:bz_by_comparison}{b}] leaves the AS rate essentially unchanged, since the AS strain fills the channel core uniformly (\figref{fig:strain_maps_z0}). The AA rate, in contrast, rises by more than two orders of magnitude and overtakes the AS contribution above $\approx 2.8$~GHz, in agreement with the AA nodal structure of \figref{fig:strain_maps_z0}. In this AA-dominated regime, the total Faraday rate exceeds the bulk value by a factor of $\sim 3.5$.

\subsection{Spin-flip rates in the Voigt configuration}\label{sec:voigt}

\begin{figure}[tb]
    \centering
    \includegraphics[width=1.0\linewidth]{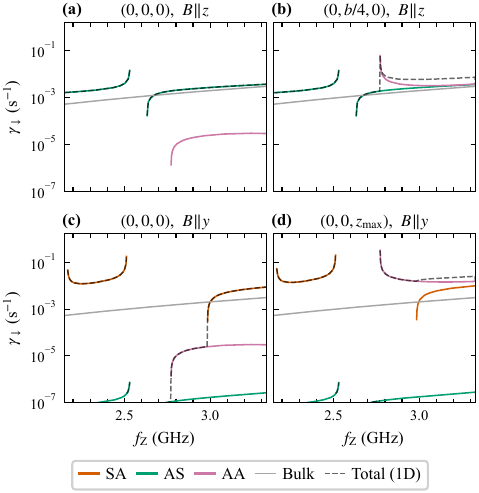}
    \caption{Spin-flip rate $\gamma_{\doa}$ as a function of Zeeman frequency $f_Z$, resolved by symmetry branches. Panels (a,b) show the Faraday configuration ($B \parallel z$) at $(0,0,0)$ and $(0,b/4,0)$ respectively. Panels (c,d) show the Voigt configuration ($B \parallel y$) at $(0,0,0)$ and $(0,0,z_{\max})$, respectively. Gray lines show the bulk reference rate. Dashed black lines show the total 1D waveguide rate. All panels at $T=4 \mr{K}$.}
    \label{fig:bz_by_comparison}
\end{figure}

\asubfigsref{fig:bz_by_comparison}{c}{d} show $\gamma_{\doa}$ versus
$f_Z$ in the Voigt configuration ($B \parallel y$), at the channel
center and at the vertical displacement $(0,0,z_{\max})$.

At the channel center [\asubfigref{fig:bz_by_comparison}{c}], the SA mode dominates $\gamma_{\doa}$ across most of the band gap, and the AS contribution is strongly suppressed. This reversal of the Faraday hierarchy follows from the strain-field projection of Section~\ref{sec:Strain_profiles} and matches the strain profiles in \figref{fig:strain_maps_z0}. The SS-only protected window of Section~\ref{sec:dispersion} persists and is effectively wider than in the Faraday case, because the AS mode also couples only weakly to the spin for $B \parallel y$. The AA branch enters this effectively widened window at $\sim 2.77$~GHz and the SA branch re-enters at $\sim 2.98$~GHz. Away from the van Hove singularities, the AA contribution sets the total rate and suppresses it by a factor of about $87$ relative to the bulk value.

Displacing the QD to $(0,0,z_{\max})$ [\asubfigref{fig:bz_by_comparison}{d}] activates the AA contribution sharply, since $\eps_{xy}$ becomes non-zero at $z_{\max}$ and couples to $B \parallel y$ through the projection of Section~\ref{sec:Strain_profiles} (see \figref{fig:strain_maps_zmax}). The SA branches dominate the rate across most of the band gap and enhance it by a factor of about $10$ relative to the bulk value. The van Hove singularities of the SA branches are analyzed in
Section~\ref{sec:powerlaw}.


\subsection{Power-law behavior at the van Hove singularities}\label{sec:powerlaw}

\begin{figure}[tb]
    \centering
    \includegraphics[width=1.0\linewidth]{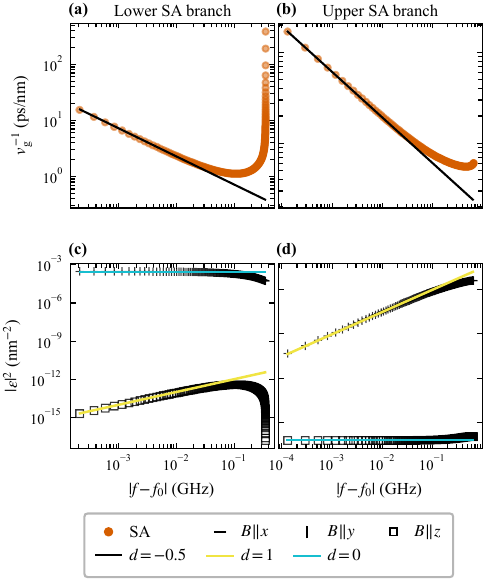}
    \caption{Power-law decomposition of the spectral density for the two SA branches at $(0,0,0)$. Panels (a,c) and (b,d) correspond to the lower and upper SA branch, with $f_0 = 2.513$~GHz and $f_0 = 2.984$~GHz, respectively. (a,b) Inverse group velocity $v_{\mr{g}}^{-1}$ (orange circles), equal to the 1D DoS per branch, together with the reference power law $|f-f_0|^{-1/2}$ ($d=-0.5$, black solid line). (c,d) Squared strain coupling $|\eps|^2$ of the spin-flip-active components for each field orientation: $B\parallel x$ (horizontal ticks), $B\parallel y$ (vertical ticks), and $B\parallel z$ (open squares). Reference power laws are shown as solid lines for $d = 1$ (yellow) and $d = 0$ (cyan).}
    \label{fig:powerlaw_SA}
\end{figure}

The sharp features at the BZ center and edges in \figref{fig:bz_by_comparison} arise from competing power-law behaviors of the DoS and of the strain coupling. At every such point, $v_{\mr{g}} \to 0$ and the parabolic dispersion forces the DoS to diverge as $|f - f_0|^{-1/2}$ [\asubfigsref{fig:powerlaw_SA}{a}{b}], where $f_0=\omega_0/(2\pi)$ denotes the frequency at the singularity. The squared strain coupling follows a power law $|\eps|^2 \propto |f - f_0|^{d}$ [\asubfigsref{fig:powerlaw_SA}{c}{d}], with an exponent set by mode symmetry and field orientation: $d = 0$ when the relevant strain components remain finite at $f_0$, and $d = 1$ when symmetry forces them to vanish. At the BZ center and edge, the Bloch mode reduces to a standing wave with definite parity under $x \to -x$, so strain components odd under this operation vanish at the dot position, while even components remain finite. The spectral density, therefore, scales as $|f - f_0|^{d - 1/2}$, diverging for $d = 0$ and vanishing as
$|f - f_0|^{+1/2}$ for $d = 1$.
The dispersion curves approach the zone center or edge as two branches of an avoided crossing (see Fig.~\ref{fig:dispersion}); one of them has $d=0$ and the other one $d=1$, following their opposite symmetry with respect to the $yz$ mirror plane at these characteristic $k$ points. 

This behavior applies to every branch (see, for example, the AS branch in \asubfigref{fig:bz_by_comparison}{a}). We illustrate it for the two SA branches at the channel center in \afigref{fig:powerlaw_SA} by decomposing the spectral density into its DoS and strain contributions. For the lower SA branch, the two Voigt orientations behave identically and give $d = 0$ while Faraday gives $d = 1$: the spectral density therefore scales as $|f - f_0|^{-1/2}$ in Voigt, producing a divergent peak at $f_0 = 2.513$ GHz, and as $|f - f_0|^{+1/2}$ in Faraday, suppressing the contribution at the zone center. For the upper SA branch, the behavior reverses, with Faraday giving $d = 0$ and Voigt giving
$d = 1$: the divergence appears at $f_0 = 2.984$~GHz in Faraday and is
suppressed in Voigt.
\par


\subsection{Non-Markovian dynamics at finite detuning}\label{sec:Result_detuning}

\begin{figure}[tb]
    \centering
    \includegraphics[width=1.0\linewidth]{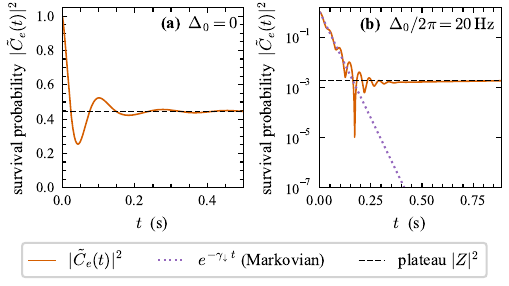}
    \caption{Survival probability of the upper Zeeman level for the lower SA BZ center at (0,0,0), Voigt configuration $\vB \parallel \hat{x}$ ($\mathcal{A} = 69.5\,\mr{s}^{-3/2}$, $T = 4$~K), at two detunings: (a) $\Delta_0 = 0$ (linear scale) and (b) $\Delta_0/2\pi = 20$~Hz (logarithmic scale). Orange solid: the exact non-Markovian survival probability. Purple dotted: the Markovian golden-rule decay. Black dashed: the trapped bound-state fraction $|Z|^2$.}
    \label{fig:Detuning}
\end{figure}

We now evaluate the Weisskopf--Wigner solution of Section~\ref{sec:ww} for the lower SA BZ center at (0,0,0) in the Voigt configuration ($\vB \parallel \hat{x}$), with $\mathcal{A} = 69.5\,\mr{s}^{-3/2}$ at $T = 4$~K, corresponding to the van Hove scale $D_0^{*}/2\pi \approx 5.77$~Hz. \afigref{fig:Detuning} shows the exact survival probability $|\tilde C_e(t)|^2$ from \eqnref{eq:ww_decomposition} at the singularity and at a small detuning. In the latter case, it is compared with the Markovian prediction $e^{-\gamma_{\doa}t}$.

At the singularity [\asubfigref{fig:Detuning}{a}] the Markovian description fails entirely: the golden-rule rate diverges, while the exact population relaxes through damped oscillations onto the bound-state plateau $|Z|^2 = 4/9$ and never decays exponentially. The non-Markovian treatment is therefore indispensable at small detuning, $\Delta_0 \lesssim D_0^{*}$, where the spin dynamics are qualitatively different from any rate description. Already at $\Delta_0/2\pi = 20$~Hz [\asubfigref{fig:Detuning}{b}], corresponding to $\delta \approx 3.47$, the dynamics develop a clean exponential stage at the rate $\Gamma' = 38.4\,\mr{s}^{-1}$, within $1.41\%$ of the golden-rule value $\gamma_{\doa}$, consistent with the leading term of \eqnref{eq:ww_gr_expansion}. The decay extends over nearly three decades before the pole--cut interference of \eqnref{eq:ww_result} appears as oscillations at the difference frequency $\Omega - \Delta_0 = w_0^2$, and the population finally settles on the reduced plateau $|Z|^2 \approx 1.84\times10^{-3}$.

The exponential stage lengthens and the plateau falls as the detuning grows: at $\Delta_0/2\pi = 200$~Hz ($\delta \approx 34.7$), the branch-cut integrand develops a Lorentzian resonance centered near $h = \Delta_0$, with full width at half maximum $\Gamma'/2\pi = 1.96$~Hz. Its Fourier transform is an exponential decay that tracks the golden rule over more than seven decades, with a fractional rate difference of $1.50\times10^{-5}$, before the population settles on the plateau $|Z|^2 \approx 2.30\times10^{-9}$.
Thus, the relaxation dynamics is practically indistinguishable from exponential already at this very small offset from the singularity.

The convergence to the golden rule is quantified in Appendix~\ref{sec:app-gr_convergence}: the fractional deviation collapses as $\delta^{-3}$ beyond the van Hove scale, so at the $1\%$ level, the non-Markovian window extends only to $\Delta_0/2\pi \approx 22.5$~Hz.

\section{Discussion and conclusions}
\label{sec:conclusions}

We have formulated a theoretical description of spin relaxation in structured phononic reservoirs based on a combination of Markovian and non-Markovian open-system theory and finite-element computations for strain profiles. The theory covers both the regular cases, where the Fermi golden rule applies, and the non-Markovian dynamics near van Hove singularities of phonon dispersion. In the latter case, we have derived a power-law dependence of the DoS and strain amplitudes and the special slow-mode selection rules unknown in bulk, from which we found universal non-exponential dynamics due to polaronic dressing of the spin into the slow acoustic mode.

Using this theory, 
we computed the phonon-induced spin-flip rate for a single electron in a self-assembled InGaAs quantum dot embedded in a W1m phononic crystal waveguide, resolved by guided-mode symmetry class as a function of magnetic-field orientation and dot position. 
Typically, the Markovian description is applicable and in most cases the relaxation is faster than in bulk at the same magnetic field. There are, however, ranges of magnetic fields for which the transition is not resonant to any of the effectively coupled modes, and spin relaxation is suppressed by many orders of magnitude. The system can be tuned to this regime by magnetic field orientation or magnitude, the latter obviously being a more practical in-situ tuning parameter. In fact, the required minor relative modulation of the Zeeman frequency can be easily achieved by tuning the $g$-factor with an electric field, which enables fast electric control of spin relaxation if the QD is additionally embedded in a field-effect structure.
In addition, the inhomogeneous mode profiles and symmetry-induced nodal planes of the strain fields lead to position dependence of the couplings, which results in strong spatial inhomogeneity of the relaxation rates.

The range of fields around the DoS singularities, where the non-Markovian dynamics is manifested, turns out to be extremely narrow, at least for the system explicitly studied here. 
This sharpness has a direct experimental consequence. Tuning the Zeeman frequency to within $20$~Hz of a van Hove singularity requires fixing the magnetic field to within a nano-Tesla, several orders of magnitude below the fluctuations of the nuclear Overhauser field in GaAs, whose rms is on the millitesla scale \cite{Hung_2013}. Therefore, any relaxation curve measured in a standard repetitive (temporal-ensemble) experiment averages over a distribution of detunings (inhomogeneous hyperfine-induced broadening) that is vastly wider than the non-Markovian window. As a consequence, the bound-state physics at the singularity is not directly accessible in this material system, except for single-shot measurements or when efficient decoupling or nuclear cooling schemes are applied. Our analysis shows that beyond $\sim 200$~Hz detuning from each van Hove singularity, the Markovian description is exact to $1.50\times10^{-5}$ or better, which puts the Born--Markov treatment on a firm quantitative footing.

Our results show that spins coupled to structured phononic reservoirs manifest relaxation dynamics much richer than those in plain bulk. This, on the one hand, paves the way for efficient magnetic or electric control of spin relaxation in phononic structures. On the other hand, the critical dependence of the spin-flip rates on the QD position, as well as on the orientation and magnitude of external fields makes these findings important for designing quantum devices. Let us note, finally, that our theoretical approach is also applicable to other spin systems, including spin-carrying defects in diamond, hBN, and other solid-state systems, where similar effects will appear in spin relaxation, governed by the particular spin-strain couplings in those systems. Similar dynamics can also be expected for phonon-assisted hyperfine spin relaxation, to which our theory can be extended, although here phonons merely close the energy gap, which may result in relaxed selection rules and mode selectivity.

\begin{acknowledgments}
This project was supported by the Polish National Science Centre (Narodowe Centrum Nauki, NCN) under grant no. 2023/50/A/ST3/00511. Claude (Anthropic, Opus, Sonnet and Fable models from version 4.5 to 5.0) was used to check and extend derivations: the author supplied an outline of the intended derivation and the author's own derivation, and accepted or rejected each output after analytic re-derivation. Claude was also used to refine the code for post-processing and plotting the finite-element data, and for citation checks and language editing. Gemini (Google) was used for literature search. The authors independently verified all calculations, literature references, and code, and assume full responsibility for the scientific accuracy of the manuscript.

\end{acknowledgments}


\appendix

\section{Eigenmode expansion and quantization}
\label{sec:app-quantize}

In this Appendix we present the details of the quantization procedure based on the confined eigenmodes of the phononic structure. 

The displacement field is expanded as
\begin{equation}\label{eq:mode_expansion}
    \bm{u}(\rr)=  \sum_{\lck} A_{\lk} \bm{u}^{\lk}(\rr),
\end{equation}
with the orthonormality condition 
\begin{equation*}
   \langle\bm{u}^{\lk}|\bm{u}^{\lambda' k'}\rangle =
   \delta_{\lambda\lambda'}\delta_{kk'}.
\end{equation*}

Following the standard quantization procedure, the generalized coordinate $A_{\lk}$ in \eqnref{eq:mode_expansion} is expressed in terms of the phonon annihilation ($b_{\lk}$) and creation operators ($b^{\dagger}_{\lnk}$) as
\begin{equation*}
    A_{\lk} = \sqrt{\frac{\hbar}{2M\omega_{\lk}}}(b_{\lck} + b_{\lcnk}^{\dagger}).
\end{equation*}
The quantized displacement is then
 \begin{equation*}
    \bm{u}(\rr) = \sum_{\lck} \sqrt{\frac{\hbar}{2M\omega_{\lk}}}\bm{u}^{\lk}(\rr)(b_{\lck}+b_{\lcnk}^{\dagger}),
\end{equation*}
where the sum over $k$ runs over both positive and negative values. 
%
The strain is similarly expanded as
\begin{align*}
    \eps_{ij}(\rr) &= \frac{1}{2} \left[\partial_i u_j(\rr)+\partial_j u_i(\rr) \right]
\\ 
    &= \sum_{\lck} \sqrt{\frac{\hbar}{2M\omega_{\lk}}}(b_{\lck}+b_{\lcnk}^{\dagger})\eps_{ij}^{\lk}(\rr),
\end{align*}
where eigenmode strain $\eps_{ij}^{\lk}(\rr)$ is given by Eq.~\eqref{eq:eps-symmetry}. Substituting the strain expansion into $H_{\mr{sph}}(\rr)$ from \eqnref{eq:H_sph} and projecting onto the QD ground state at position $\rr_0$, using
$|\Psi_0(\rr)|^2 \approx \delta(\rr-\rr_0) $ (that is, the QD is treated as a point-like object compared to the phonon wavelength),
\begin{align*}
H_{\mr{sph}} &= \int d^3r \, \Psi_0^*(\rr) \, H_{\mr{sph}}(\rr) \Psi_0(\rr) \\      &= \sum_{\lck} \alpha_j(\lck) \sigma_j 
(b_{\lck} + b_{\lcnk}^{\dagger}),
\end{align*}
with the coupling constant
\begin{equation*}
\alpha_j(\lambda, k) = \frac{1}{2} \sum_{i} 
\sqrt{\frac{\hbar}{2 M \omega_{\lambda k}}}\, \eta\, \mu_{\mr{B}} B_i\, 
\eps_{ij}^{\lambda k}(\rr_0).
\end{equation*}
Setting $B_z=B, B_x,B_y=0$ and omitting the term containing $\sigma_z$ (see Sec.~\ref{sec:theory-formalism} of the main text), we get the spin-phonon Hamiltonian in the form of Eq.~\eqref{eq:H_sph_alpha}, with $\alpha(\lck) = \alpha_x(\lck) - i \alpha_y(\lck)$. 

\section{Spin-flip rates in bulk}
\label{sec:app-bulk}


In this Appendix we derive the spin-phonon spectral density for a bulk semiconductor, following the same formalism as in Sec.~\ref{sec:theory}. In bulk, the displacement field is expanded in plane waves

\begin{equation}
\bm{u}(\rr) = \sum_{\vlk} \sqrt{\frac{\hbar}{2\rho V \omega_{\vlk}}} e^{i \vk \cdot \rr} \hat{e}(\vlck)  ( b_{\vlck} + b^{\dagger}_{\lambda, \bm{-k}}),
\end{equation}
where $V$ is the volume, $\rho$ is the mass density of the bulk crystal, and $\hat{e}(\vlck)$ is the polarization vector. The coupling constant takes the form

\begin{align}\label{eq:alpha_bulk}
\alpha_j(\vlck)  &= \frac{i}{4}   \sum_{i} \sqrt{\frac{\hbar}{2\rho V \omega_{\vlk}}} \mu_{\mr{B}}  \eta \notag\\ &\times [k_j  \hat{e}_i(\vlck) +k_i \hat{e}_j(\vlck)] B_i  F(\vk),
\end{align}
where $F(\vk)=\exp[-(\ell k/2)^2]$ is the Gaussian form factor associated with the finite QD size $\ell$. For the acoustic phonons of interest, $\ell \approx 4$~nm is much smaller than the phonon wavelength and the supercell of the waveguide structure, so that $\ell k \ll 1$ and $F\approx1$. This corresponds to the point-like approximation $|\Psi_0(\rr)|^2 \approx \delta(\rr-\rr_0)$ used in Sec.~\ref{sec:theory}.

We choose the $z$-axis along the magnetic field, so that $B_i=B_z$. Using $\alpha(\lck) = \alpha_x(\lck) - i \alpha_y(\lck)$ and the spectral density from \eqnref{eq:zeta_1D}, we convert the sum over $\vk$ to an integral in spherical coordinates. To evaluate the sum over polarizations, we use the completeness relation for the transverse and longitudinal branches. The resulting bulk spectral density is

\begin{align}\label{eq:zeta_bulk}
\zeta(\omega) &=\frac{1}{\pi^2} \cdot \frac{1}{80}    \frac{ \eta^2 \mu_{\mr{B}}^2 }{\hbar \rho} B_z^2 \, \omega^3 \notag \\  & \times \Biggl[ \frac{2}{3}\cdot\frac{1}{c_{\mr{l}}^5}\Bigl|F\left(\frac{\omega}{c_{\mr{l}}}\right)\Bigr|^2+  \frac{1}{c_{\mr{t}}^5}\Bigl|F\left(\frac{\omega}{c_{\mr{t}}}\right)\Bigr|^2  \Biggr],
\end{align}
where $c_{\mr{l}}$ and $c_{\mr{t}}$ denote the longitudinal and transverse sound velocities, respectively. The relaxation rates are determined by \eqref{eq:gamma_down}.

\section{Non-Markovian dynamics}
\label{sec:app-non-markov}

In this Appendix we present the details of the derivations for the non-Markovian dynamics near van Hove singularities and link the dynamics of the system to the analytic structure of $\tilde C_e(s)$. This is divided into three parts: First, we discuss the solution in the case of Zeeman energy resonant with one of the dispersion branches (within the 1D band). Next, we analyze in quantitative detail the convergence to the Fermi golden rule from the transient dynamics. Finally, we discuss the off-resonant case, when the transition energy lies in the gap between the branches. 

\subsection{Resonant case}
\label{sec:app-non-markov-reson}

Projecting the Schr\"odinger equation onto the two states appearing in \eqnref{eq:ww_ansatz} gives
\begin{align}
\frac{d\tilde C_e}{dt} &= \frac{1}{i\hbar}\sum_{\lck}\sqrt{n_{\lk}+1}\;\alpha(\lck)\, e^{i\Delta\omega_{\lk} t}\,\tilde C_{\lk}(t),
\label{eq:ww_eom_e}\\
\frac{d\tilde C_{\lk}}{dt} &= \frac{1}{i\hbar}\sqrt{n_{\lk}+1}\;\alpha^*(\lck)\, e^{-i\Delta\omega_{\lk} t}\,\tilde C_e(t),
\label{eq:ww_eom_k}
\end{align}
where $\Delta\omega_{\lk} = \omz - \omega_{\lk}$. Integrating \eqnref{eq:ww_eom_k} formally and substituting the result into \eqnref{eq:ww_eom_e} yields the equation
\begin{align}
\frac{d\tilde C_e(t)}{dt} ={}& -\frac{1}{\hbar^2}\sum_{\lck}\bigl(n_{\lk}+1\bigr)\,\bigl|\alpha(\lck)\bigr|^2 \notag\\
&\times\int_0^t dt'\, e^{i\Delta\omega_{\lk}(t-t')}\,\tilde C_e(t').
\label{eq:ww_memory}
\end{align}
The occupations enter the memory kernel only through the factor $(n_{\lk}+1)$. The dynamics is dominated by modes very close to the van Hove singularity, whose Zeeman angular frequency we denote by $\omega_0$, so we can replace $n_{\lk} = n_{\mr{B}}(\omega_0)$ assuming thermal equilibrium.
Eq.~\eqref{eq:ww_memory} is solved via the Laplace transform. For the initial condition $\tilde{C}_e(0)=1$, the Laplace-domain solution $\tilde C_e(s)$ of Eq.~\eqref{eq:ww_memory} is
\begin{equation}
\tilde C_e(s) = \frac{1}{s + E(s)},
\label{eq:ww_resolvent-app}
\end{equation}
where the self-energy is the Laplace transform of the memory kernel and can be expressed as
\begin{equation}
E(s) = \int_0^\infty d\omega\, \frac{\Xi(\omega)}{s - i(\omz - \omega)}.
\label{eq:ww_selfenergy-app}
\end{equation}
Here we define the Bose-weighted spectral density
\begin{equation*}
\Xi(\omega) = \bigl[n_{\mr{B}}(\omega_0)+1\bigr]\,\zeta(\omega),
\end{equation*}
with $\zeta(\omega)$ from \eqnref{eq:zeta_1D}. 

Near a dispersion extremum at angular frequency $\omega_0 = 2\pi f_0$, the dispersion is parabolic, $\omega_{\lk} = \omega_0 + \beta(k-k_0)^2$, so the DoS in \eqnref{eq:zeta_final} reduces to $\lvert dk/d\omega\rvert = \tfrac{1}{2}[\beta(\omega-\omega_0)]^{-1/2}$ and the spectral density vanishes below $\omega_0$. For $\omega \ge \omega_0$,
\begin{equation}
\Xi(\omega) = \mathcal{A}\,(\omega-\omega_0)^{-1/2},
\label{eq:ww_edge-app}
\end{equation}
with the thermal prefactor $\mathcal{A}$ given by Eq.~\eqref{eq:A-prefactor},
written for a dispersion minimum ($\beta>0$), with the strain components taken in the frame of Sec.~\ref{sec:theory-formalism}, whose $z$-axis lies along $\vB$. The description at a dispersion maximum follows by reflection of the spectrum about $\omega_0$. In that case, the detuning entering all subsequent expressions is $\Delta_0 = \omega_0 - \omz$, so that positive $\Delta_0$ again places the Zeeman frequency inside the band. The opposite case of negative detuning, in which the Zeeman frequency lies inside the band gap, is analyzed in Sec.~\ref{sec:app-negdet} of this Appendix.
In the parabolic approximation, the frequency integral in \eqnref{eq:ww_selfenergy-app} is elementary,
\begin{equation}
E(s) = \frac{-i\pi\mathcal{A}}{\sqrt{-\Delta_0 - is}},
\qquad \Delta_0 = \omz - \omega_0,
\label{eq:ww_E-app}
\end{equation}
where the principal branch of the root has to be chosen, $\mr{Re}\sqrt{-\Delta_0 - is} \ge 0$. Substituting into Eq.~\eqref{eq:ww_resolvent-app} yields Eq.~\eqref{eq:ww_resolvent} of the main text.

\begin{figure}[tb]
\centering
\colorlet{axiscolor}{black}
\colorlet{branchcut}{brown!70!orange}
\colorlet{polecolor}{red!75!black}
\colorlet{sheetcolor}{black!70}
\colorlet{bromwichcolor}{teal!80!black}
\colorlet{jordancolor}{blue!70!black}
\colorlet{axislegcolor}{violet!70!black}

\begin{tikzpicture}[
    scale=0.9,
    axis/.style={->, thick, >=Stealth, draw=axiscolor},
    bromwich_style/.style={thick, bromwichcolor, postaction={decorate, decoration={markings,
        mark=at position 0.55 with {\arrow[thick]{>}}}}},
    jordan_style/.style={thick, jordancolor, postaction={decorate, decoration={markings,
        mark=at position 0.55 with {\arrow[thick]{>}}}}},
    jordan2_style/.style={thick, jordancolor, dashed, postaction={decorate, decoration={markings,
        mark=at position 0.55 with {\arrow[thick]{>}}}}},
    leg_style/.style={thick, axislegcolor, postaction={decorate, decoration={markings,
        mark=at position 0.5 with {\arrow[thick]{>}}}}},
    leg2_style/.style={thick, axislegcolor, dashed, postaction={decorate, decoration={markings,
        mark=at position 0.5 with {\arrow[thick]{>}}}}},
    loop_style/.style={thick, axislegcolor, postaction={decorate, decoration={markings,
        mark=at position 0.35 with {\arrow[thick]{>}}}}},
    loop2_style/.style={thick, axislegcolor, dashed}
]
\draw[axis] (-4.1,0) -- (4.41,0) node[right, font=\small, text=black] {$\mr{Re}\,z$};
\draw[axis] (0,-2.5) -- (0,4.31) node[above, font=\small, text=black] {$\mr{Im}\,z$};

\draw[branchcut, very thick, decorate, decoration={zigzag, amplitude=1.5pt, segment length=5pt}] (-3.9,0) -- (0,0);
\filldraw[branchcut] (0,0) circle (2.4pt);
\node[branchcut, font=\small, anchor=north] at (-2.5,-0.32) {branch cut};

\filldraw[polecolor] (1.9,0) circle (2.4pt);
\node[polecolor, font=\small, anchor=south] at (1.9,0.34) {$z_0=w_0^2$};

\draw[sheetcolor, thick] (-1.7,1.15) circle (2.4pt);
\node[sheetcolor, font=\small, anchor=south] at (-1.7,1.34) {$z_2=w_2^2$};
\draw[sheetcolor, thick] (-1.7,-1.15) circle (2.4pt);
\node[sheetcolor, font=\small, anchor=north] at (-1.7,-1.25) {$z_1=w_1^2$};

\draw[bromwich_style] (-3.55,-1.9) -- (3.55,-1.9);
\draw[jordan_style] (3.55,-1.9) arc[start angle=-28.2, end angle=90, radius=4.0];
\draw[leg_style] (0.12,4.0) -- (0.12,0.54);
\draw[loop_style] (0.12,0.54) arc[start angle=77.4, end angle=-180, radius=0.55];
\draw[loop2_style] (-0.55,0) arc[start angle=180, end angle=102.6, radius=0.55];
\draw[leg2_style] (-0.12,0.54) -- (-0.12,4.0);
\draw[jordan2_style] (0,4.0) arc[start angle=90, end angle=180, radius=4.0];
\draw[jordan_style] (-4.0,0) arc[start angle=180, end angle=208.2, radius=4.0];

\node[bromwichcolor, font=\small, align=center] at (1.5,-2.55)
    {Bromwich line\\ $\mr{Im}\,z=-c$};
\node[jordancolor, font=\small] at (2.55,2.55) {$C_R$};
\node[axislegcolor, font=\small, anchor=west] at (0.3,3.1) {$C_{\mathrm{rem}}$};
\node[axislegcolor, font=\small, anchor=north west] at (0.45,-0.45) {$C_\delta$};

\end{tikzpicture}
\caption{Deformed contour used to isolate the second-sheet pole, drawn in the variable $z$. Solid lines lie on the first Riemann sheet, dashed lines on the second. The first-sheet pole $z_0=w_0^2$ is shown as a filled circle and the second-sheet poles by open circles.}
\label{fig:ww_contour}
\end{figure}

The amplitude of the upper spin state is recovered by the inverse Laplace transform of $\tilde C_e(s)$, for which we use Mellin's formula. For this, it is convenient to introduce the transformed variable $z = -\Delta_0 - is$. In this variable, the Bromwich integral reads
\begin{equation}
\tilde C_e(t) = \frac{e^{i\Delta_0 t}}{2\pi i}\int_{-\infty-ic}^{\infty-ic} dz\;
\frac{\sqrt{z}e^{izt}}{\sqrt{z}(z+\Delta_0) - \pi\mathcal{A}},
\label{eq:ww_bromwich}
\end{equation}
along a horizontal Bromwich line $\mr{Im}\,z = -c$ placed to the bottom of every singularity. The contour is deformed to expose the second-sheet pole, as shown in Fig.~\ref{fig:ww_contour}. From the Bromwich line the large arc $C_R$ reaches the positive imaginary axis on the first sheet. Along the positive imaginary axis the contour descends on the first sheet, circles the branch point at $z=0$ along the small circle $C_\delta$ of vanishing radius, and ascends on the second sheet. The circle runs three quarters on the first sheet and the last quarter on the second after crossing the cut. The large arc then continues on the second sheet from the imaginary axis back to the branch cut, where it returns to the first sheet and closes onto the Bromwich line. The contributions of $C_R$ vanish as $R\to\infty$ by Jordan's lemma, and those of $C_\delta$ vanish with its radius. The two imaginary-axis segments $C_{\mathrm{rem}}$ lie on different sheets and combine into $I_{\mathrm{rem}}(t)$, Eq.~\eqref{eq:I-rem}. Of the second-sheet poles $z_{1,2}=w_{1,2}^2$, with $z_2=z_1^*$ (open circles in Fig.~\ref{fig:ww_contour}), $z_2$ is enclosed and $z_1$ is not.

By writing $w = z^{1/2}$, with $\mr{Re}\,w \ge 0$ on the first sheet, the poles of the integrand of Eq.~\eqref{eq:ww_bromwich} can be found from the cubic equation
\begin{equation}
w^3 + \Delta_0 w - \pi\mathcal{A} = 0.
\label{eq:ww_cubic}
\end{equation}
Note that this equation admits the scaling $w = (\pi\mathcal{A})^{1/3}v$ and $\Delta_0 = D_0^{*}\,\delta$, under which it collapses to $v^3 + \delta v - 1 = 0$.
For $\Delta_0 \ge 0$, \eqnref{eq:ww_cubic} has a single real root $w_0 > 0$ and a
complex-conjugate pair $w_{1,2} = -\tfrac{1}{2}w_0 \pm i\chi$ with
$\chi = \sqrt{\Delta_0 + \tfrac{3}{4}w_0^2}$. Only the real root $z_0=w_0^2$ lies on the first sheet and  contributes to the integral, corresponding to an undamped bound-state contribution. The value of $z_0$ starts from $(\pi\mathcal{A})^{2/3} = D_0^{*}$ for $\Delta_0 = 0$ and reaches the asymptotic behavior $z_0 \approx (\pi\mathcal{A}/\Delta_0)^2$ as $\Delta_0\to\infty$. 
The first-sheet pole at $z_0=w_0^2$ gives, via the residue theorem, the bound-state term of Eq.~\eqref{eq:ww_decomposition} with 
$\Omega=\Delta_0+z_0=\Delta_0 + w_0^2$ and 
$Z = [(d/dz)(z+\Delta_0-\pi\mathcal{A}/\sqrt{z})]_{z=w_0^2}^{-1}$, 
where we substitute  $\pi\mathcal{A} = w_0(w_0^2 + \Delta_0)$.

Besides the bound-state pole at $z_0$ the deformed contour picks up contributions from the pole at $z_2 = w_2^2= -\frac{1}{2}w_0^2 - \Delta_0 +iw_0\chi$ and from the segment along the positive part of the imaginary axis, first on the first and then on the second sheet, denoted by $C_{\text{rem}}$ in Fig.~\ref{fig:ww_contour}. 
The pole yields an exponentially decaying contribution. As $\Delta_0$ grows, the imaginary part of $z_2$ behaves asymptotically as $\pi\mathcal{A}/\sqrt{\Delta_0}=\gamma_{\downarrow}/2$.
The integral along $C_{\text{rem}}$ is calculated by parametrizing the path along the positive imaginary semi-axis by $z=iy$, $y>0$, and combining the contributions from the first and second sheet. This gives the non-exponential remainder,
\begin{equation}\label{eq:I-rem}
 I_{\text{rem}}(t) = 
\mathcal{A}e^{i\Delta_0 t-i\pi/4}
\int_0^\infty dy
\frac{\sqrt{y}e^{-yt}}
{y\left(\Delta_0+i y\right)^2+i\pi^2\mathcal{A}^2}.
\end{equation}
In this way, the deformed contour yields the  survival amplitude in the form of Eq.~\eqref{eq:ww_decomposition} of the main text, rigorously splitting the dynamics into bound (non-decaying), exponential and power-law contributions.

As noted in the main text, $Z'\to 1$ when $\Delta_0\to\infty$, so that the solution is fully dominated by the exponentially decaying term in this limit, which restores the Fermi golden rule far away from the singularity. 
More generally, the integrand of Eq.~\eqref{eq:I-rem} behaves as $-i\sqrt{y}/(\pi\mathcal{A})^2$, which leads, via Watson's lemma, to the long-time asymptotics of this term
\begin{equation}
\label{eq:I-rem-longt}
 I_{\text{rem}}(t) \approx 
\frac{e^{i\Delta_0 t-3i\pi/4}}{2\pi^{3/2}\mathcal{A}t^{3/2}}.
\end{equation}
The long-time spin amplitude is therefore
\begin{equation}
\tilde C_e(t\to\infty) \simeq \frac{2w_0^2}{3w_0^2 + \Delta_0}\,e^{i\Omega t}
+ \frac{e^{-i3\pi/4}}{2\pi^{3/2}\,\mathcal{A}\,t^{3/2}}\,e^{i\Delta_0 t}.
\label{eq:ww_amplitude}
\end{equation}
The two terms beat at the difference frequency $\Omega - \Delta_0 = w_0^2$, and the survival probability of the upper Zeeman level follows in the form of Eq.~\eqref{eq:ww_result}.
The endpoint expansion that yields \eqnref{eq:I-rem-longt}, carried to next order in $1/t$, shows when this algebraic description itself becomes reliable: its correction term reaches the order of unity at the crossover time \begin{equation}
    t_c \sim \frac{\Delta_0^2}{\pi^2\mathcal{A}^2}.
    \label{eq:ww_tc}
\end{equation}
Because $t_c$ depends on the detuning only through $\Delta_0^2$, this expression applies for either sign of $\Delta_0$. Watson's lemma itself is valid once $t$ exceeds the inverse of the distance from the band edge to the nearest remaining singularity of the remainder integrand, which is the bound-state pole at $z_0 = w_0^2$ (the second-sheet poles lie farther, at $|z_{1,2}| = w_0^2 + \Delta_0$), i.e.\ for $t \gg 1/w_0^2$, which reduces to the same parametric scale as $t_c$ once $\Delta_0 \gg D_0^{*}$.

\subsection{Convergence to the Fermi golden rule}
\label{sec:app-gr_convergence}

In this section we quantify how  the decay rate associated with the second-sheet pole approaches the golden-rule rate $\gamma_\doa$ as the reduced detuning $\delta$ grows. This analysis allows us also to determine the width of the non-Markovian window around the singularity, quoted in Section~\ref{sec:Result_detuning}. 

The occupation decay rate is $\Gamma' = 2\,\mr{Im}\,z_2 = 2w_0\chi$, with $\Gamma'\to\gamma_\doa$ as $\Delta_0\to \infty$. The Fermi golden rule is recovered with a cubically small correction. Expanding $w_0$ and $\chi$ for $\Delta_0 \gg D_0^{*}$ gives
\begin{equation}
\frac{\Gamma'}{\gamma_\doa} = 1 - \frac{5}{8}\,u + \frac{231}{128}\,u^2 + \mathcal{O}(u^3),
\qquad u = \left(\frac{D_0^{*}}{\Delta_0}\right)^{3},
\label{eq:ww_gr_expansion}
\end{equation}
where $\gamma_\doa = 2\pi\mathcal{A}/\sqrt{\Delta_0} = 2\pi\,\Xi(\omz)$ is the Markovian rate of \eqnref{eq:gamma_down} evaluated in the parabolic approximation for the spectral density. The deviation from the golden rule therefore collapses as $\Delta_0^{-3}$ once the detuning exceeds the van Hove scale, and $D_0^{*}$ provides a sharp definition of the non-Markovian window.

\begin{figure}[tb]
    \centering
    \includegraphics[width=1.0\linewidth]{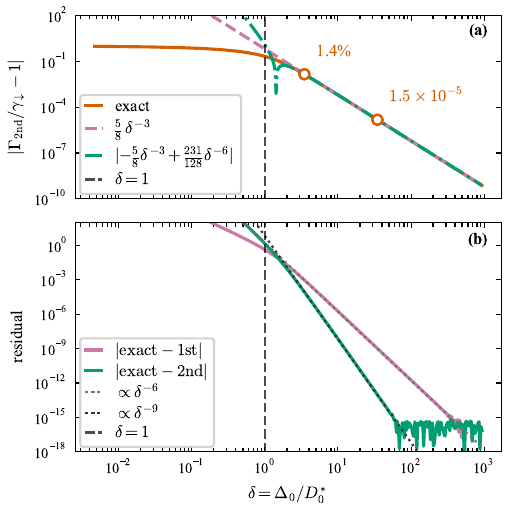}
    \caption{Convergence of the transient decay rate to the golden rule, versus the reduced detuning $\delta$. (a) Fractional deviation of the exact second-sheet rate from the Markovian golden-rule rate: orange solid line marks the exact curve; pink dashed line marks the leading term of \eqnref{eq:ww_gr_expansion}; green dash-dotted, the magnitude of the two-term sum. Open orange circles mark the two detunings quoted in the text. (b) Residuals of the series truncated at first (pink solid line) and second (green solid line) order. Gray and black dotted lines are the reference power laws $\delta^{-6}$ and $\delta^{-9}$, respectively, and the floor at $\sim10^{-15}$ is double-precision roundoff. Black dashed vertical line in both panels: $\delta = 1$.}
    \label{fig:gr_correction}
\end{figure}

The comparison is shown in \afigref{fig:gr_correction}. The exact fractional deviation $|\Gamma'/\gamma_\doa - 1|$ collapses onto the leading term $\tfrac{5}{8}\,\delta^{-3}$ of \eqnref{eq:ww_gr_expansion} immediately beyond the van Hove scale, and the residuals of the truncated series follow the expected $\delta^{-6}$ and $\delta^{-9}$ power laws. The cubic collapse makes the non-Markovian window remarkably sharp: the deviation is $1.41\%$ at $20$~Hz and drops to $1.50\times10^{-5}$ at $200$~Hz. Adopting a $1\%$ deviation from the golden rule as the boundary, the window extends to $\delta \approx 3.91$, that is, to $\Delta_0/2\pi \approx 22.5$~Hz. Measured against the $1.18$~GHz band gap, it spans a fractional frequency range of $1.9\times10^{-8}$.

\subsection{Off-resonant case: transition in the gap}
\label{sec:app-negdet}

When the Zeeman frequency $\omz$ falls below the phonon band edge ($\Delta_0 = \omz - \omega_0 < 0$), no resonant phonon mode exists and the Fermi golden rule gives zero decay rate. The spin cannot relax by emitting a real phonon. It does, however, hybridize with the evanescent tail of the continuum, forming a spin-phonon bound state that protects part of the population from relaxation. The rest leaks into propagating modes through a non-exponential transient and disperses as phonons. Deeper into the gap, the overlap with the continuum shrinks and the protection grows: the bound-state population rises from $4/9$ at the band edge toward unity at large $|\Delta_0|$.

The self-energy \eqnref{eq:ww_E-app}, the cubic \eqnref{eq:ww_cubic}, and the resolvent \eqnref{eq:ww_resolvent-app} hold for either sign of $\Delta_0$. The cubic retains a single positive real root $w_0$, Vieta's relations give $w_1 + w_2 = -w_0$ and $w_1 w_2 = \pi\mathcal{A}/w_0 > 0$, so $\mr{Re}\,w_{1,2} < 0$ throughout and neither root satisfies the principal-branch condition. The bound-state pole $z_0 = w_0^2$ [Eq.~\eqref{eq:ww_decomposition}] therefore remains the sole singularity on the physical (first) Riemann sheet (the sheet reached by analytic continuation of $\tilde C_e(s)$ from the convergent half-plane), with $\Omega = \Delta_0 + w_0^2 = w_0^2 - |\Delta_0| > 0$ at every negative detuning (from the cubic, $w_0^3 > |\Delta_0|\,w_0$). The bound state stays real and undamped throughout the gap.

The bound-state weight follows from \eqnref{eq:ww_Z} without modification:
\begin{equation}
    Z(\Delta_0) = \frac{2w_0^2}{3w_0^2 + \Delta_0} = \frac{2w_0^3}{2w_0^3 + \pi\mathcal{A}},
    \label{eq:neg_Z}
\end{equation}
the fraction of the spin amplitude carried by the bound state, giving a locked population $P(\infty) = Z^2$ [\eqnref{eq:ww_result}]. $Z$ is positive, analytic through $\Delta_0 = 0$, and increases monotonically from $2/3$ at the band edge toward $1$ deep in the gap, with exact values $Z = 2n/(2n+1)$ at $w_0^3/\pi\mathcal{A} = n$ (\tabref{tab:negdet}). At deep detuning, $w_0 \simeq |\Delta_0|^{1/2}$ and
\begin{equation}
    1 - Z \simeq \frac{\pi\mathcal{A}}{2|\Delta_0|^{3/2}}, \qquad
    \Omega \simeq \frac{\pi\mathcal{A}}{\sqrt{|\Delta_0|}},
    \label{eq:neg_deep}
\end{equation}
confirmed by second-order perturbation theory in the coupling to the continuum. The leaked weight vanishes as $|\Delta_0|^{-3/2}$ and the bound-state frequency $\Omega$ approaches zero: the dynamics reduce to static level repulsion with no relaxation.

The amplitude decomposes into the bound-state pole, the second-sheet pole, and a remainder integral, the three terms of \eqnref{eq:ww_decomposition} with $I_{\text{rem}}(t)$ from \eqnref{eq:I-rem}, as long as the second-sheet pole $z_2$ lies inside the deformed contour, that is for $\Delta_0 > \Delta_0^{\dagger} \approx -0.794\,D_0^{*}$. The square root in the self-energy is non-analytic at $z = 0$ (the band edge), and the remainder integral along the positive imaginary axis captures the weight that disperses into the phonon continuum. For $\Delta_0 > \Delta_0^{\dagger}$, including the band edge and shallow negative detuning, the second-sheet pole $z_2$ produces the transient exponential decay at rate $\Gamma'$ (Section~\ref{sec:app-gr_convergence}). For $\Delta_0 < \Delta_0^{\dagger}$ the pole $z_2$ crosses the positive imaginary axis and lies outside the contour, the $Z'$ term is absent, the released population disperses entirely through the algebraic remainder, reaching the plateau $Z^2$ through interference oscillations at frequency $w_0^2$ [\eqnref{eq:ww_result}] without a preceding exponential ring-down.


The long-time tail has the same form at every detuning. The behavior of the remainder integrand near $y = 0$ (the point where the self-energy diverges as $y^{-1/2}$) is controlled by the denominator's value $\pi^2\mathcal{A}^2$ there, which is independent of $\Delta_0$. Watson's lemma, an asymptotic expansion of the integral dominated by this endpoint singularity, then returns the algebraic tail of \eqnref{eq:ww_amplitude} with the same $t^{-3/2}$ power law. Its amplitude and phase are set by the square-root van Hove singularity at the band edge, not by the detuning. The population approaches
$P(\infty) = Z^2$, following \eqnref{eq:ww_result}. For $|\Delta_0| \gg D_0^{*}$ the crossover time is again $t_c$ of \eqnref{eq:ww_tc}, the same expression as at positive detuning to relative accuracy $(D_0^{*}/|\Delta_0|)^3$. Near the band edge the sign of $\Delta_0$ changes $t_c$ at order one. At deep detuning this scale equals $1/w_1^2$, the inverse distance of the shallowest virtual level $z_1 = w_1^2$ from the band edge (with $w_1$ the real negative root of smaller magnitude). Deep detuning therefore postpones the onset of the algebraic regime without modifying the tail itself.

The physical picture inverts relative to positive detuning. For $\Delta_0 > 0$ the bound state traps a small, asymptotically vanishing fraction under golden-rule decay, for $\Delta_0 < 0$ it is the dominant feature, holding a population $Z^2$ that grows from $4/9$ at the band edge toward $1$ at deep detuning (\tabref{tab:negdet}). Tuning the Zeeman frequency outside the band protects at least $4/9$ of the spin population against spin flip, and nearly all of it once $|\Delta_0| \gg D_0^{*}$. All expressions are continuous through $\Delta_0 = 0$, since $w_0(\Delta_0)$ is analytic there.

The remaining structure in \tabref{tab:negdet} belongs to the analytic continuation, not to the dynamics. The remaining pair $w_{1,2}$ (conjugate above the exceptional point, both real and negative below it) merges into a double root at $\Delta_0^{\mr{EP}} = -3(\pi\mathcal{A}/2)^{2/3} \approx -1.890\,D_0^{*}$ and separates below it as two real, negative virtual levels: poles of $\tilde C_e(s)$ on the unphysical (second) Riemann sheet, with zero width and no counterpart on the physical sheet. At $\Delta_0^{\dagger} \approx -0.794\,D_0^{*}$, these second-sheet poles cross the imaginary axis, and $z_2$ passes through the remainder segment of the contour. Both events rearrange the analytic continuation but leave $\Omega$, $Z$, and $|\tilde C_e(t)|^2$ analytic, with no transition in the dynamics. The table's exact anchors $Z = 2n/(2n+1)$ at $n = 1,\,2,\,4$ coincide with $\Delta_0 = 0$, $\Delta_0^{\dagger}$, and $\Delta_0^{\mr{EP}}$.

\begin{table}[h]
    \centering
    \begin{tabular}{r|l|c|c|c}
        $\delta$ & root structure & $w_0$ & $Z$ & $P(\infty)$\\
        \hline
        $0$ & band edge & $1$ & $2/3$ & $4/9$\\
        $-0.794$ & submersion & $1.260$ & $4/5$ & $16/25$\\
        $-1.2$ & complex pair & $1.386$ & $0.842$ & $0.709$\\
        $-1.890$ & coalescence & $1.587$ & $8/9$ & $64/81$\\
        $-4$ & real pair & $2.115$ & $0.950$ & $0.902$\\
        $-100$ & deep gap & $10.005$ & $0.9995$ & $0.9990$\\
    \end{tabular}
    \caption{Bound-state weight and asymptotic population for $\Delta_0 \le 0$, in reduced units $\pi\mathcal{A} = 1$ ($w_0$ in units of $(\pi\mathcal{A})^{1/3}$, $\delta = \Delta_0/D_0^{*}$ with $D_0^{*}$ of \eqref{eq:ww_D0}). $Z$ follows from \eqnref{eq:neg_Z} and $P(\infty) = Z^2$. The root-structure column labels the arrangement of the cubic's roots: at $\delta^{\dagger} = -2^{-1/3}$ the second-sheet poles cross the imaginary axis, at $\delta^{\mr{EP}} = -3\cdot2^{-2/3}$ the pair coalesces. Both are events of the analytic continuation and leave $|\tilde C_e(t)|^2$ analytic.}
    \label{tab:negdet}
\end{table}


%




\bibliography{references}


\end{document}